\documentclass[aip,reprint,amsmath,amssymb]{revtex4-1}
\usepackage[T1]{fontenc}
\usepackage[latin9]{inputenc}
\usepackage{color}
\usepackage{babel,lipsum}
\usepackage{amsthm,float}
\usepackage{amsmath,physics}
\usepackage{amssymb}
\usepackage{calrsfs}
\usepackage{graphicx}

\usepackage[unicode=true,pdfusetitle, bookmarks=true,bookmarksnumbered=false,bookmarksopen=false,  breaklinks=false,pdfborder={0 0 0},backref=false,colorlinks=false] {hyperref}
\hypersetup{ colorlinks,linkcolor=myurlcolor,citecolor=myurlcolor,urlcolor=myurlcolor}
\usepackage{txfonts}
\usepackage{colortbl}\definecolor{myurlcolor}{rgb}{0,0,0.7}
\usepackage{soul}
\usepackage{qcircuit}
\DeclareMathAlphabet{\pazocal}{OMS}{zplm}{m}{n}

\makeatletter
\@ifundefined{textcolor}{}
{%
 \definecolor{BLACK}{gray}{0}
 \definecolor{WHITE}{gray}{1}
 \definecolor{RED}{rgb}{1,0,0}
 \definecolor{GREEN}{rgb}{0,1,0}
 \definecolor{darkgreen}{rgb}{0,0.8,0}
 \definecolor{BLUE}{rgb}{0,0,1}
 \definecolor{CYAN}{cmyk}{1,0,0,0}
 \definecolor{MAGENTA}{cmyk}{0,1,0,0}
 \definecolor{YELLOW}{cmyk}{0,0,1,0}
 }

\theoremstyle{plain}
\newtheorem{thm}{\protect\theoremname}
\theoremstyle{plain}
\newtheorem{prop}[thm]{\protect\propositionname}
\ifx\proof\undefined
\newenvironment{proof}[1][\protect\proofname]{\par
\normalfont\topsep6\p@\@plus6\p@\relax
\trivlist
\itemindent\parindent
\item[\hskip\labelsep
\scshape
#1]\ignorespaces
}{%
\endtrivlist\@endpefalse
}
\providecommand{\proofname}{Proof}
\fi
\theoremstyle{plain}

\newtheorem{theorem}[thm]{\protect\theoremname}
\newtheorem{defi}[thm]{\protect\definitionname}

\newtheorem*{theorem*}{(Theorem}
\newtheorem*{prop*}{(Proposition}
\newtheorem*{lemma*}{(Lemma}
\newtheorem*{defi*}{(Definition}
\newtheorem*{corollary*}{Corollary}

\providecommand{\lemmaname}{Lemma}
\providecommand{\theoremname}{Theorem}
\providecommand{\definitionname}{Definition}
\providecommand{\propositionname}{Proposition}
\providecommand{\corollaryname}{Corollary}

\begin{document}

\title{A local-realistic theory for fermions}

\author{Nicetu Tibau Vidal}
\affiliation{Clarendon Laboratory, University of Oxford, Parks Road, Oxford OX1 3PU, United Kingdom}

\author{Vlatko Vedral}
\affiliation{Clarendon Laboratory, University of Oxford, Parks Road, Oxford OX1 3PU, United Kingdom}
\affiliation{Centre for Quantum Technologies, National University of Singapore, 3 Science Drive 2, Singapore 117543}
\affiliation{Department of Physics, National University of Singapore, 2 Science Drive 3, Singapore 117542}
\author{Chiara Marletto}
\affiliation{Clarendon Laboratory, University of Oxford, Parks Road, Oxford OX1 3PU, United Kingdom}

\begin{abstract}
    We propose a local model for general fermionic systems, which we express in the Heisenberg picture. To this end, we shall use a recently proposed formalism, the so-called "Raymond-Robichaud" construction, which allows one to construct an explicitly local model for any dynamical theory that satisfies no-signalling, in terms of equivalence classes of transformations that can be attached to each individual subsystem. By following the rigorous use of the parity superselection rule for fermions, we show how this construction removes the usual difficulties that fermionic systems display in regard to the definition of local states and local transformations.
\end{abstract}

\maketitle
\section{Introduction}
\label{sec:intro}

Since the well-known works of Bell \cite{Bell1,Bell2} quantum theory has been labelled as 'non-local'. This terminology arises from the fact that quantum systems have been proven, both theoretically and experimentally, to violate Bell's inequalities. As a consequence of this fact, Bell's theorem implies that quantum theory cannot admit a local hidden variable model in terms of a real-valued stochastic theory -- hence the term non-local. \\

However, quantum theory has long been known to be local in a deeper, dynamical sense. First, it satisfies the so-called "no-signalling theorem``; moreover, it satisfies the principle of no-action at a distance, as formulated by Einstein \cite{EIN}. \\

The fact that the principle of no-action at a distance is satisfied is most explicit in the Heisenberg picture formulation. That is the case in both non-relativistic and relativistic quantum theory \cite{Gottesman,Deutsch}. Specifically, for a system of $N$ qubits, it is possible to define elements of reality attached to each qubit, which fully specify the dynamics of the composite system of $N$ qubits jointly. We refer to these elements of reality as 'descriptors' \cite{Deutsch}. Such descriptors are not real-valued elements of a stochastic hidden-variable theory, precisely as Bell's theorem requires. They are q-numbers \cite{qnnumbers}, i.e., elements of a non-commutative algebra, which generate the algebra of local observables of each qubit. One can recover the dynamics of the system of all qubits by simply considering the set of all local descriptors of each qubit.\\

This relation between no-signalling and no-action at a distance, which emerges in quantum theory, is not accidental. Recent work \cite{Raymond} proved formally and in a general way that every no-signalling theory with reversible dynamics admits a so-called local-realistic model, which Raymond-Robichaud \cite{Raymond} defines as a model where: (i) each system has a mathematical object that describes its real factual situation; (ii) given the objects (say, $a$ and $b$) fully describing one of two disjoint subsystems $A$ \& $B$, the objects describing the global system $AB$ are entirely determined by the ordered pair $(a,b)$; (iii) a dynamical transformation that affects system $A$ only, can only change the object $a$ describing $A$, but cannot change the object $b$ describing $B$. The objects $a$ and $b$ are called local elements of reality, or the ontic states of the subsystems $A$ and $B$. Furthermore, the authors give a constructive method to define such local elements of reality of any theory that is no-signalling and dynamically reversible - which we shall refer to as the {equivalence class formalism}. Raymond-Robichaud \cite{Raymond2} uses this {formalism} to find the local elements of reality in qubit networks; such local elements of reality can be represented in terms of the standard local descriptors of qubits, i.e. the generators of the algebra of observables \cite{Charles}. \\

In this paper, we shall apply these recent results to fermionic quantum field theory. We will show how it admits local elements of reality as q-numbers by explicitly constructing them following the {equivalence class formalism}. As one would naturally expect, we will demonstrate that the familiar generators of the fermionic algebra can be used to represent the local elements of reality. Our work provides the essential ground to understand better the foundations of fermionic phenomena, such as the fermionic Aharonov-Bohm effect \cite{ABeffect}. Moreover, it helps to ease further the conceptual difficulties with the locality of fermionic quantum field theory and seek a unified description of classical and quantum theories. \\

The structure of the paper is as follows. We first define fermionic systems (section \ref{sec:fermions}) and explain the formal notion of local realism in fermionic field theory (section \ref{sec:localrealism}). Then, we use the structure of fermionic theory to derive the fermionic descriptors using the {equivalence class formalism} (section \ref{sec:descriptors}). Finally, we discuss the results, commenting on the subtleties of the formalism and suggesting generalisations of our results (section \ref{sec:discussion}).

 \section{Fermionic theory}
\label{sec:fermions}

Fermions can be regarded as a type of quantum system. In this work, we shall adopt a general information-theoretic approach to represent them, following previous works \cite{Friis,Nicetu}. This perspective adheres to the position of studying locality in fermionic systems in terms of modes \cite{Friis13}, not particles \cite{Schliemann01, Eckert02} . Within this approach, a fermionic system consists of a set of modes $I$ and for each mode $i\in I$ there are fermionic creation and annihilation operators $\hat{f}_i, \hat{f}_i^\dagger$. The ground state $\ket{\Omega}$ is defined by the requirement that $\hat{f}_i \ket{\Omega}=0$. The characteristic fermionic operator algebra is given by the creation and annihilation operators of all modes $i\in I$ obeying the following commutation relations: 

\begin{eqnarray}
    \{\hat{f}_i,\hat{f}_j\}=\hat{0} \qquad \qquad \{\hat{f}_i,\hat{f}_j^\dagger\}=\delta_{i j} \hat{\mathbb{I}}
\end{eqnarray}

We focus throughout the paper on systems with a finite set of modes for simplicity. In the  Schr\"odinger picture, the allowed physical states are spanned by the states obtained by acting with the creation operators on the ground state $\ket{\Omega}$, allowing linear combinations of states with the same parity only. With parity, we mean the parity of the number of fermions in the state. This restriction to allow only superpositions of states with the same parity is the parity superselection rule. Remarkably, this rule must be introduced for fermions to satisfy the no-signalling principle.\\

As an example, the density operator $\rho=\frac{1}{2}\left(\ketbra{\Omega}+\ketbra{1}+\ketbra{\Omega}{1}+\ketbra{1}{\Omega}\right)$ is not allowed to physically exist, otherwise we could have faster than light communication between two parties. Formally, one can state the parity {superselection rule} as requiring that any physically allowed density operator $\rho$ has to commute with the parity operator $\mathbb{P}$. $\mathbb{P}$ has as eigenstates of eigenvalue $1$($-1$) the states with an even (odd) number of fermions. This {parity superselection rule} is long known \cite{Wigner} to restrict the set of physically allowed density operators in a fermionic system. However, it also restricts the set of unitaries and observables that one can apply to the fermionic system. It has been shown recently \cite{Nicetu} that physical unitaries and observables also need to commute with the parity operator $\mathbb{P}$ in order not to be able to signal.\\  

More concretely, in a two-mode system, unitaries and observables with the form $\hat{f}_1+\hat{f}_1^\dagger$ are not physically allowed.  It follows from the {parity superselection} rule that if $\hat{U}_I$ is a {parity superselection rule} unitary that involves modes in the set $I$, and we consider a mode $j\notin I$, then the action of $\hat{U}_I$ on $\hat{f}_j$ leaves it invariant: $\hat{U}_I^\dagger \hat{f}_j \hat{U}_I=\hat{f}_j$. This result is a direct implication of $\hat{U}_I$ commuting with the parity operator. This fact is crucial for when we analyse in the Heisenberg picture the local mode structure. We will see that a local unitary in $A$ does not affect the elements of reality of a disjoint region $B$. However, before going into detail about descriptors and the Heisenberg picture, let us first set the basis of what we mean by locality. We adhere to the position that locality is a structural requirement for a theory. We believe that the definition provided by Raymond-Robichaud \cite{Raymond} formalises perfectly this notion of locality at a structural level. 

 \section{Local realism}
\label{sec:localrealism}

After providing a general definition of local-realistic physical theories, Raymond-Robichaud \cite{Raymond} proves that every physical theory satisfying the no-signalling principle and having reversible dynamics is a local-realistic theory. In this section, we apply this theorem to fermions. We first introduce the notion of a realistic theory, then clarify the notion of locality that we use. Finally, we show how fermionic theory fits into the {construction} of the {equivalence class formalism}. \\   

\subsection{Realism}
\label{subsec:realism}

{To define a realistic theory, we use a triad of sets $(\pazocal{P},\pazocal{R}, \pazocal{T})$ on which we impose the following structures.}\\

The set $\pazocal{P}$ is the set of the elements of the physical theory that describe the phenomenal properties of the theory - i.e., those that are empirically accessible. In other words, an element $\rho\in \pazocal{P}$ details all the properties that can be observed of a specific configuration of a physical system. For fermionic theories, this is the set of density operators $\rho$ that satisfy the {parity superselection rule}. We call the set $\pazocal{P}$ the phenomenal state space and its elements phenomenal states since they give us the observational properties of the system. \\

The set $\pazocal{T}$ is the set of operations that describe the allowed physical transformations. In the fermionic case, this corresponds to the set of {parity superselected} unitary operators. The set of transformations for fermions forms a group {under composition $\circ$}. This fact gives reversible dynamics in the theory. The transformations {group} $\pazocal{T}$ acts on the phenomenal state space $\pazocal{P}$, defining a group action. The group action in the fermionic case is given by $\rho'= \hat{U} \cdot  \rho \cdot  \hat{U}^\dagger$, where $\hat{U}\in\pazocal{T}$ and $\rho,\rho'\in \pazocal{P}$.\\

The set $\pazocal{R}$ is the set of states of a physical system describing {\sl all} properties that entirely specify the configuration of that physical system, rather than merely describing the observational properties of it. We call the set $\pazocal{R}$ the ontic state space. Since the term ontic appears more often in the literature \cite{ontic} than the terms real or noumenal, we will be using it throughout this work. The group of operations $\pazocal{T}$ also acts on $\pazocal{R}$, but not necessarily with the same action as in $\pazocal{P}$. We denote the action of an element $U\in \pazocal{T}$ to an element $r\in \pazocal{R}$ as $r'=U\star r$. We will see in section \ref{sec:descriptors} how the descriptors are a representation of these ontic states.\\

Let us consider a familiar example to exemplify the meaning of the ontic states. Consider a qubit composite system with the singlet state $\ket{\psi}=\frac{1}{\sqrt{2}} \left(\ket{01}-\ket{10}\right)$, and we apply a local rotation in one of the qubits. We know that the Schr\"odinger state $\ket{\psi}$ will remain unchanged. One can imagine in this case that the phenomenal state is $\ket{\psi}$ since it gives the observational properties of the system. However, we can consider that the fact that we have applied a rotation is relevant for the constitutive properties of the system, even though such rotation does not produce any observational effect. In that case, we could consider that the realised rotation is part of our ontic state.\\ 

The phenomenal state must be completely characterised by the ontic state of the system: there must be an epimorphism $\varphi$ from $\pazocal{R}$ to $\pazocal{P}$. In other words, there has to be a mapping $\varphi: \pazocal{R} \to \pazocal{P}$ where for every $\rho \in \pazocal{P}$ exists a $r\in \pazocal{R}$ such that $\varphi(r)=\rho$. Moreover, $\varphi$ has to be faithful with respect to the two actions of the group of operations $\pazocal{T}$ onto the sets $\pazocal{R}$ and $\pazocal{P}$. In the fermionic case, for all $\hat{U}\in \pazocal{T} , r\in \pazocal{R}$ then $\varphi(\hat{U}\star r)=\hat{U} \cdot \varphi(r) \cdot \hat{U}^\dagger$.\\

Thus, if we want to see that the fermionic theory is a realistic theory, we have to find a set $\pazocal{R}$, an appropriate group action $\star$ and an epimorphism $\varphi$ that satisfy the conditions above. One could choose $\pazocal{R}$ as being $\pazocal{P}$, where the action of $\pazocal{T}$ is the same as in $\pazocal{P}$ and the epimorphism is the trivial identification $\varphi(\rho)=\rho$. However, this trivial identification may not follow the conditions to be considered a local-realistic theory. In order to show that fermionic quantum field theory is a local-realistic theory, we need to find a suitable ontic state space for a fermionic theory that satisfies the locality properties described below.\\  

\subsection{Locality}
\label{subsec:locality}

The notion of locality presented \cite{Raymond} does not refer to position in any way. It refers to the structure of subsystems \cite{Chiribella}. So, when we talk about locality, we refer to being localised in a subsystem of the global system in consideration. \\

As we comment in section \ref{sec:fermions}, we take the fermionic mode as the primary subsystem of a fermionic system. More concretely, if we have a physical system of $N$ modes $\pazocal{N}=\{1, \dots, N\}$, we consider that the systems where the set of modes is a subset of $\pazocal{N}$ are the physical subsystems of the theory. \\

The notion of locality refers to the relation between different fermionic modes of the set $\pazocal{N}$. We need to specify which systems our states refer to and how our transformations act in terms of mathematical structure. We will call an object local to a subsystem $A$ if it only refers to or acts on that subsystem $A$.  We are also interested in restricting objects that live or act in the global system to the particular subsystems. \\

In the phenomenal state space $\pazocal{P}$, this so-called projection map $\pi^{\pazocal{P}}$ is given by the partial trace. If we have a phenomenal state of the whole system of $N$ modes $\rho$, we say that the object that gives all the phenomenal properties of the system within the subsystem of modes $\pazocal{M}=\{r_1,\dots,r_M\}\subseteq \pazocal{N}$ is given by the fermionic partial trace of $\rho$ over the set of modes $\pazocal{N} \backslash \pazocal{M}$ denoted by $\rho_{r_1,\dots,r_M}=\tr_{\pazocal{N}\backslash \pazocal{M}} (\rho)$. Indeed $\rho_{r_1,\dots,r_M}$ is a valid phenomenal state, since it is {a parity superselected} density operator. Moreover, it can be considered to be local to the subset of modes $\pazocal{M}=\{r_1,\dots,r_M\}$ since $\rho_{r_1,\dots,r_M}$ can be given by combining creation and annihilation operators of the $\pazocal{M}$ modes alone.\\

For operations, it is also necessary to have a notion of an operation being local to a subset of modes $\pazocal{M}$. For fermions, an operation that acts on the global system of modes $\pazocal{N}$,  $\hat{U}\in \pazocal{T}$,  is local to the subsystem given by the subset of modes $\pazocal{M}$ iff $\hat{U}$ can be given by a combination of creation and annihilation operators of the modes in $\pazocal{M}$ alone. \\

\subsection{Fermions in the {equivalence class formalism}}
\label{subse:ferminsRR}

We now analyse what we are required to obtain in order to find a local-realistic theory for fermions. First, we need a realistic theory; thus, we need an ontic state space with an action of the transformation space and the epimorphism $\varphi$ to the fermionic phenomenal state space. In order for a realistic theory to be local for a lattice of systems, we need to be able to define two operations in the ontic state space $\pazocal{R}$. First, we need a projection map $\pi^{\pazocal{R}}_S: \pazocal{R} \to \pazocal{R}$. The projection map sends global ontic states to local ontic states. We require that the ontic projections and the phenomenal projections act in parallel. It is required that for any $\rho$ being the phenomenal state of the whole system of fermionic modes $\pazocal{N}$, then for the ontic state(s) $R\in \pazocal{R}$ such that $\varphi(R)=\rho$, then $\varphi(\pi^{\pazocal{R}}_S(R))=\rho_S=\pi^{\pazocal{P}}_S(\rho)$, where $S$ is any subset of modes of $\pazocal{N}$. So we want a projection to subsystems that is faithful under the $\varphi$ epimorphism. These properties give us a notion of ontic states belonging to a subsystem.  \\

Last and crucially, we want a structure where knowing the ontic states of the parts, we know the ontic state of the whole. More specifically, we want to be able to define a map $\odot : \pazocal{R}_A \times \pazocal{R}_B \to \pazocal{R}_{AB}$. Consider a phenomenal state $\rho\in \pazocal{P}$ of the whole system of modes, such that it has an underlying ontic state $R\in \pazocal{R}$ given by $\varphi(R)=\rho$. Also, consider that then we split the global system of modes in two disjoint pieces by choosing two disjoint subsets of modes $A$ and $B$, such that $A \cup B=\pazocal{N}$ and $A\cap B=\emptyset$. Then we can define ontic states $R_A=\pi^{\pazocal{R}}_A(R)$, $R_B=\pi^{\pazocal{R}}_B(R)$ by the previously defined projection operation. Now, we require that we can assign $R_A \odot R_B=R_{A\cup B}=R$ uniquely. Observe that this is the condition of separability in the Einstein locality principle \cite{philosopher}. Moreover, notice that with the usual understanding that density operators are the phenomenal states and the ontic states of the system, such join product cannot be defined. Given an entangled state $\rho$ for the disjoint systems $A$ and $B$, then one can obtain $\pi^{\pazocal{P}}_A(\rho)=\tr_B(\rho)=\rho_A$ and $\pi^{\pazocal{P}}_B(\rho)=\tr_A(\rho)=\rho_B$, where the projection map is the partial trace. But from the un-entangled product state $\rho'=\rho_A \wedge \rho_B$ using the same map we also get to $\pi^{\pazocal{P}}_A(\rho')=\tr_B (\rho')=\rho'_A=\rho_A$ and $\pi^{\pazocal{P}}_B(\rho')=\tr_A(\rho')=\rho'_B=\rho_B$. Thus, we cannot define a join product $\rho_A \odot \rho_B=\rho$ and $\rho_A \odot \rho_B=\rho'$ since $\rho\neq \rho'$. Thus, the usual realistic view of fermionic theory is not local realistic. However, this does not stop us from discussing if it is possible to find a different ontic state space where the local realistic structure manifests. We see in section \ref{sec:descriptors} how using the constructive proof of the {equivalence class formalism theorem}, we find such structure. \\ 

 In the {equivalence class formalism}, an operational no-signalling theory needs a phenomenal state space $\pazocal{P}$, a set of transformations $\pazocal{T}$ with an action $\cdot$ on $\pazocal{P}$. Moreover, it needs a lattice of systems with associated faithful projectors $\pi^{\pazocal{P}}_S$ and that the no-signalling condition is respected. We can understand the condition of $\pazocal{T}$ having a group structure and giving a group action on $\pazocal{P}$ as that the dynamics of the theory are reversible. This reversibility is the case in fermionic theory since the set of transformations $\pazocal{T}$ corresponds to the set of fermionic {parity superselected} unitaries for a given complete set of modes. Such unitaries set, of course, conforms a group with the usual product. The phenomenal state space $\pazocal{P}$ for fermionic theory is, as mentioned before, the set of {parity superselected} density operators for the given total set of modes. Given $\hat{U}\in \pazocal{T}$, $\rho \in \pazocal{P}$ the group action is defined as $\hat{U}\cdot \rho \cdot \hat{U}^\dagger$. Furthermore, the partial trace over modes in the complete set of modes gives the projections of the system lattice in the fermionic case. If we have a total set of modes $\pazocal{N}$, we can define any subsystem by any subset $\pazocal{M}\subseteq \pazocal{N}$ and we obtain $\pi^{\pazocal{P}}_{\pazocal{M}}(-)=\tr_{\pazocal{N} \backslash \pazocal{M}} (-)$. \\

Finally, we express the no-signalling principle presented in the {equivalence class formalism} as: given a system with a bipartition of disjoint subsystems $A$ and $B$, any operation local in $A$ $U_A\in \pazocal{T}$, any operation local in $B$ $V_B\in \pazocal{T}$ and any phenomenal state of the global system $AB$ $\rho\in \pazocal{P}$ the following equation is satisfied:

\begin{eqnarray}
   \pi^{\pazocal{P}}_A\left(\left(U_A V_B\right) \cdot \rho \right)=U_A\cdot \left(\pi^{\pazocal{P}}_A\left(\rho\right)\right)    
\end{eqnarray}

which in fermionic theory corresponds to the usual no-signalling principle equation given by 

\begin{eqnarray}
  &\tr_B\left(\hat{U}_A \cdot \hat{V}_B \cdot \rho \cdot  \hat{V}_B^\dagger \cdot \hat{U}_A^\dagger\right)=\hat{U}_A \cdot \tr_B\left(\rho\right) \cdot \hat{U}_A^\dagger  \Leftrightarrow \nonumber \\ &\Leftrightarrow \tr_B\left(\hat{V}_B \cdot \rho \cdot \hat{V}_B^\dagger\right)=\tr_B\left(\rho\right)
\end{eqnarray}

Therefore, since the above no-signalling condition is satisfied for the fermionic system that considers the parity { superselection rule} for both states and unitaries \cite{Nicetu}, all the conditions of the {equivalence class formalism} are satisfied. Therefore, we can find with their construction the ontic states that dote fermionic theory of the structure of local realism.\\ 

The proof that any operational no-signalling theory is a local-realistic theory is constructive \cite{Raymond}. The authors construct for a general operational theory with a group of transformations $\pazocal{T}$, a lattice of systems, a phenomenal state space $\pazocal{P}$, with the associated group action $\cdot$ and projection operators $\pi^{\pazocal{P}}_A$ the ontic state space $\pazocal{R}$, the ontic group action $\star$, the epimorphism $\varphi$, the ontic projection operators $\pi^{\pazocal{R}}_A$ and the join product $\odot$. They do so by defining an equivalence relation $\sim_A$ that states that two transformations $U,V$ are equivalent in a subsystem $A$ $U\sim_A V$ if and only if exists a transformation $W_{\bar{A}}$ local on the subsystem $\bar{A}$,  complementary to $A$, such that $U= W_{\bar{A}} V$. The equivalence classes $[U]_A$ under this equivalence are the ontic states of the subsystem $A$, forming an ontic state space $\pazocal{R}$ with all the desired properties. Some of them are: 

\begin{eqnarray*}
    V \star [U]_A= [VU]_A \qquad \pi^{\pazocal{R}}_A\left([U]_{A\cup B}\right)=[U]_A \end{eqnarray*} \begin{eqnarray}  [U]_A \odot [U]_B =[U]_{A\cup B} \qquad W_A \star [V]_B=[V]_B 
\end{eqnarray}

where $A$ and $B$ are systems, with $A\cap B=\emptyset$ and  $U,V,W_A$ are transformations with $W_A$ being local in $A$. \\

In order to find these equivalence classes in the fermionic setting, we utilise the equivalence between the descriptors and the {equivalence class formalism} in qubit networks \cite{Bedard}. The paper aims to generalise it to the fermionic case and use the excellent properties and physical intuition that the descriptor picture provides to unveil the set of ontic states that give the fermionic theory a local realistic structure.\\

\section{Descriptors}
\label{sec:descriptors}

As we have explored in section \ref{sec:localrealism}, in order to provide a fermionic theory with a local realistic structure for an orthonormal set of modes, we need to find a set of ontic states $\pazocal{R}$ with proper operations $\pi^{\pazocal{R}}_A$, $\varphi$ and $\odot$. In order to construct these objects presented in the {equivalence class formalism}, we utilise the descriptor formalism presented by Deutsch \& Hayden \cite{Deutsch} and then we use the result by B\'{e}dard \cite{Bedard} that establishes the equivalence between the two formalisms.\\   

 Their work focuses on qubit networks, but we will see that one can apply their ideas to fermionic systems. Here we will be applying their procedure directly to the fermionic case. Using the Heisenberg picture, the descriptor formalism consists of re-interpreting the real elements of quantum mechanics. The formalism considers pure states with unitary evolutions, {assuming that one can always increase the dimension of the Hilbert space and use} the Stinespring dilation theorem \cite{Nielsen} {to treat any quantum channel as a unitary evolution}. Thus, from here on, we consider the Heisenberg state being pure unless stated otherwise. \\

The derivation of descriptors starts with the fact that in quantum theory, the expectation value of an observable $\hat{O}$ of a system in a particular $\rho$ is given by $\tr\left(\hat{O} \cdot \rho\right)$. Now, if we apply a unitary transformation $\hat{U}$ to $\rho$, the updated data will be given by $\tr\left(\hat{O}\cdot  \hat{U} \cdot \rho \cdot  \hat{U}^\dagger\right)$. We use the Heisenberg picture. We consider that the state $\rho$ remains invariant under the transformation $\hat{U}$, but the observables $\hat{O}$ evolve under the transformation given by the unitary $\hat{U}$ as $\hat{O}'= \hat{U}^\dagger \cdot \hat{O} \cdot \hat{U}$. Since the observables of the system have evolved and $\rho$ has not, it is reasonable to assume that we could find a framework where the observables are the ones that determine the state of the system and not $\rho$. \\

In order to do so, it is helpful to try to find which operators we can track to determine any $\hat{U}^\dagger \cdot \hat{O} \cdot \hat{U}$. We start by defining that the space that we work with consists of a fermionic space of $N$ modes under the {parity superselection rule}. Therefore, the observables $\hat{O}$ are Hermitian operators that commute with the parity operator.  Using the anticommutation relations of the creation and annihilation operators, we can write any observable $\hat{O}$ as 

\begin{eqnarray}
    \hat{O}=\sum_j \alpha_j \hat{O}_1^{(j)} \cdot \dots \cdot \hat{O}_N^{(j)}
\end{eqnarray} 

where $\hat{O}_i^{(j)}$ are monomials of any degree of the creation and annihilation operators $\hat{f}_i,\hat{f}_i^\dagger$. Notice that for each $j$ in the sum the global operator $\hat{O}_1^{(j)} \cdot \dots \cdot \hat{O}_N^{(j)}$ is a monomial of even degree of $(\hat{f}_i,\hat{f}_i^\dagger)_{i=1}^N$. Now, we can see that for any $\hat{O}$ we have this decomposition and then:

\begin{eqnarray}
   \hat{U}^\dagger \cdot  \hat{O} \cdot \hat{U}=\sum_j \alpha_j \hat{U}^\dagger \cdot \left(\hat{O}_1^{(j)}\cdot  \dots \cdot \hat{O}_N^{(j)}\right) \cdot \hat{U}=\nonumber \\ =\sum_j \alpha_j \left(\hat{U}^\dagger \cdot \hat{O}_1^{(j)}\cdot \hat{U} \right)\cdot \dots\cdot\left( \hat{U}^\dagger \cdot \hat{O}_N^{(j)} \cdot \hat{U}\right)
\end{eqnarray}

Therefore, if we know how all the possible local operators $\hat{O}_i^{(j)}$ transform under $\hat{U}$, then we know how any observable $\hat{O}$ evolves under $\hat{U}$. However, we can reduce this further. Due to the fermionic algebra, it is easy to see that there are only four linearly independent monomials of $f_i,f_i^\dagger$. Thus, the transformation under $\hat{U}$ of $\hat{O}_i^{(j)}$ will be given by:

\begin{eqnarray}
     \hat{U}^\dagger \cdot \hat{O}_i^{(j)} \cdot \hat{U} = \begin{array}{l}  \hat{U}^\dagger \cdot \hat{f}_i\cdot \hat{U} \quad \text{or} \quad \hat{U}^\dagger \cdot \hat{f}_i^\dagger \cdot  \hat{U} \quad \text{or} \\ \hat{U}^\dagger \cdot \left(\hat{f}_i \cdot \hat{f}_i^\dagger\right) \cdot \hat{U} \quad \text{or} \quad \hat{U}^\dagger \cdot \left(\hat{f}_i^\dagger \cdot \hat{f}_i \right) \cdot \hat{U} \end{array}
\end{eqnarray}

If we draw the  attention to the second line, we can observe that $\hat{U}^\dagger \cdot \left(\hat{f}_i \cdot \hat{f}_i^\dagger\right) \cdot \hat{U}=\left(\hat{U}^\dagger \cdot \hat{f}_i \cdot \hat{U}\right)\cdot \left( \hat{U}^\dagger \cdot \hat{f}_i^\dagger \cdot \hat{U} \right)$ and that $\hat{U}^\dagger \cdot \left(\hat{f}_i^\dagger \cdot \hat{f}_i\right) \cdot \hat{U}=\left(\hat{U}^\dagger \cdot \hat{f}_i^\dagger \cdot \hat{U}\right)\cdot \left( \hat{U}^\dagger \cdot \hat{f}_i \cdot \hat{U} \right)$. Therefore, we can conclude that if we know the transformation under $\hat{U}$ of the $2N$ operators $\hat{f}_i$ and $\hat{f}_i^\dagger$ for $i=1\dots N$ then we know the evolution under $\hat{U}$ of any observable $\hat{O}$. However, we can even reduce this further, since it can be observed that $ \left(\hat{U}^\dagger \cdot \hat{f}_i \cdot \hat{U} \right)^\dagger=\hat{U}^\dagger \cdot \hat{f}_i^\dagger \cdot \hat{U}$. \\

This fact implies that we only need to know the evolution under $\hat{U}$ of the $N$ operators $\hat{f}_i$ in order to know the evolution under $\hat{U}$ of any observable of the $N$ mode fermionic system. We call the set $(\hat{f}_1,\dots, \hat{f}_N)$ the set of fermionic descriptors of an $N$ mode fermionic system. \\

By knowing the initial state of the system $\ketbra{\psi}$ and knowing the expression of $\left(\hat{U}^\dagger \cdot \hat{f}_1 \cdot \hat{U},\dots, \hat{U}^\dagger \cdot \hat{f}_N \cdot \hat{U} \right)$ one can recover all the expectation values of any observable in the current state of the system. In the Schrodinger picture, such phenomenal state would be given by $\hat{U} \ket{\psi}$. A few things to note is that each $\hat{U}^\dagger \cdot \hat{f}_i \cdot \hat{U}$ can be a global $N$ mode operator, so it can have contributions in its expression from other modes $j\neq i$. But at the same time, the "updated" descriptors satisfy the fermionic algebra. In the sense that if we name $\bar{\hat{f}}_i=\hat{U}^\dagger \cdot \hat{f}_i \cdot \hat{U}$ then $\{\bar{\hat{f}}_i, \bar{\hat{f}}_j\}=0$ and $\{\bar{\hat{f}}_i, \bar{\hat{f}}_j^\dagger\}=\delta_{ij} \hat{\mathbb{I}} $\\

\subsection{Descriptors as ontic states}
\label{subsec:descriptreal}

The curious reader may be wondering how these objects relate to the notion of local realism that we have introduced in section \ref{sec:localrealism}. The answer is inspired {by} the work of B\'{e}dard \cite{Bedard}. The set of descriptors $\hat{U}^\dagger \cdot (\hat{f}_1, \dots, \hat{f}_N) \cdot \hat{U}$ is a compact way to represent the equivalence classes $[\hat{U}]_i$ that according to the {equivalence class formalism} can be considered as the ontic states of the system. This process endows fermionic theory with a local-realistic structure. \\

We develop the specifics of this claim in the following lines, where we present the results that justify the connection between the fermionic descriptors that we have proposed and the {equivalence class formalism} of local-realistic theories.  

\begin{theorem}
\label{thm:1}
Given a fermionic theory with a set of modes  $I=\{i_1,\dots,i_N\}$. Given the equivalence relation on the group of {parity superselected} unitaries of the theory for each mode $i_j\in I$ given by $\hat{U}\sim_{i_j} \hat{V}$ iff $\hat{U}=\hat{W}_{I\backslash\{i_j\}} \cdot \hat{V}$, then 

\begin{eqnarray}
    \hat{U} \sim_{i_j} \hat{V} \qquad  \Longleftrightarrow \qquad \hat{U}^\dagger \cdot \hat{f}_{i_j} \cdot \hat{U}=\hat{V}^\dagger \cdot \hat{f}_{i_j} \cdot \hat{V}
    \label{eq:equiv}
\end{eqnarray}
Thus, $[\hat{U}]_{i_j}=\{\hat{V}\in \pazocal{T}| \hat{U}^\dagger \cdot \hat{f}_{i_j} \cdot \hat{U}=\hat{V}^\dagger \cdot \hat{f}_{i_j} \cdot \hat{V} \}$

\end{theorem}

With Theorem \ref{thm:1} we have obtained a direct connection between the fermionic descriptors and the ontic states of the {equivalence class formalism}. The proof follows easily from the algebraic properties of unitaries, and it is in Appendix \ref{sec:appendix}.  We can define the equivalence classes that are the local elements of reality in terms of properties satisfied by the descriptors. Moreover, we see that we can define the equivalence class that gives the local ontic state in a subsystem in terms of only the descriptors associated with that subsystem.\\  

\subsubsection{Ontic action}
\label{subsubsec:realaction}

Using the result from Theorem \ref{thm:1} we can represent the ontic fermionic states of the {equivalence class formalism} using the fermionic descriptors. The action $\star$ of the group of transformations $\pazocal{T}$ on the ontic state space $\pazocal{R}$, $W\star [U]_S$ that we have defined in section \ref{sec:localrealism}  in the representation by descriptors is given by 
\begin{eqnarray}
  W \star \left(\left(\hat{U}^\dagger \hat{f}_1 \hat{U},\dots,\hat{U}^\dagger \hat{f}_N \hat{U}\right),\ketbra{\psi_0}\right)=\nonumber \\=   \left(\left(\hat{W}^\dagger \hat{U}^\dagger \hat{f}_1 \hat{W} \hat{U},\dots,\hat{W}^\dagger \hat{U}^\dagger \hat{f}_N \hat{W} \hat{U}\right),\ketbra{\psi_0}\right) 
\end{eqnarray}

In what follows, we express the join product $\odot$, the ontic projectors $\pi_A^\pazocal{R}$ and the epimorphism $\varphi$ in respect of fermionic descriptors. Consider an $N$ mode fermionic system with set of modes $I=\{i_1,\dots,i_N\}$. We have an initial phenomenal Heisenberg state $\rho_0=\ketbra{\psi_0}$. The system undergoes a specific transformation $\hat{U}$ given as a polynomial of the operators $\hat{f}_{i_j},\hat{f}_{i_k}^\dagger$. After such transformation is applied, the pure state $\ket{\psi}=\hat{U} \ket{\psi_0}$ is the phenomenal state of the system. In regards to the descriptor picture, we have that in respect of the lattice of systems given by the canonical mode set $I$, the set of descriptors $(\hat{f}_{i_1},\dots,\hat{f}_{i_N})$ and the initial Heisenberg state $\ket{\psi_0}$ represent the initial ontic state of the system. After the transformation $\hat{U}$ is applied the ontic state of the system can be represented by $(\hat{U}^\dagger \hat{f}_{i_1} \hat{U},\dots, \hat{U}^\dagger \hat{f}_{i_N} \hat{U})$ and the Heisenberg state $\ket{\psi_0}$.\\

\subsubsection{Descriptor epimorphism}
\label{subsubsec:epimorphism}

A result that is important to understand the completeness of the descriptor picture and its relationship with Theorem \ref{thm:1}, the group of transformations $\pazocal{T}$ and its equivalence classes is the following theorem \ref{thm:1.5}. The following theorem helps to understand the definition of the epimorphism $\varphi$ for a complete set of descriptors.

\begin{theorem}
\label{thm:1.5}
 Using the complete set of descriptors $\left(\hat{\bar{f}}_1,\dots,\hat{\bar{f}}_N\right)$ is possible to find uniquely the $\hat{U}$ that acts on the $N$ mode system such that $\left(\hat{\bar{f}}_1,\dots,\hat{\bar{f}}_N\right)=\left(\hat{U}^\dagger \hat{f}_1 \hat{U},\dots,\hat{U}^\dagger \hat{f}_N \hat{U}\right)$. 
\end{theorem}

The proof of this result is entirely algebraic, but the idea is to use a decomposition of the unitary operators on a basis of the operators that act as descriptors. The complete proof is in the Appendix \ref{sec:appendix}.\\

The construction of the epimorphism $\varphi$ seems straightforward from this theorem. We choose the annihilation operators to describe the ontic state of our initial configuration, with a given Heisenberg state $\ketbra{\psi_0}$. We say that for such configuration, the associated phenomenal state is the Heisenberg state. Then if we have a different set of descriptors that is unitarily conjugated to the canonical one, one can obtain the unitary $\hat{U}$ that connects them, and associate to the new configuration the phenomenal state $    \varphi\left(\left(\hat{U}^\dagger \hat{f}_{i_1} \hat{U}, \dots, \hat{U}^\dagger \hat{f}_{i_N} \hat{U}\right), \ketbra{\psi_0}\right)=\hat{U}\ketbra{\psi_0} \hat{U}^\dagger$.\\

This definition is consistent with what one might expect since we start with a phenomenal state $\ketbra{\psi_0}$ that we use as a Heisenberg state. The mentioned Heisenberg state and a corresponding set of canonical annihilation operators give the initial ontic state. Then, a unitary evolution makes the states evolve. The action of the unitary updates the descriptors. Moreover, since we can deduce the unitary, the new phenomenal state is the unitary acting on the initial phenomenal state. So, the definition of the epimorphism is consistent with our physical intuition on how transformations should act on the ontic and phenomenal states. \\

However, this only defines the epimorphism for a total set of descriptors, so for the global states of the system. We now present how from the ontic local state $\left(\left(\hat{U}^\dagger \hat{f}_{i_1} \hat{U}\dots , \hat{U}^\dagger \hat{f}_{i_M} \hat{U}\right), \ketbra{\psi_0}\right)$ we associate the phenomenal state $\rho_M$, which is a {parity superselected} density operator for the system of $M$ modes $i_1,\dots,i_M$. The idea is the same as for the proof of theorem \ref{thm:1.5}. It is not difficult to see that $\rho_M$ can be decomposed in an operator basis $\hat{O}_{M}^{(j)}$ given by monomials of creation and annihilation operators $\hat{f}_{i_1}, \hat{f}_{i_1}^\dagger, \dots,\hat{f}_{i_M}, \hat{f}_{i_M}^\dagger$, then manipulating the expression we get: 

\begin{eqnarray}
    \rho_M=\sum_j \tr\left( \hat{O}_M^{(j)} \rho_M\right) \hat{O}_M^{(j)}= \sum_j \tr\left(\hat{O}_M^{(j)} \rho \right) \hat{O}_M^{(j)}=\nonumber \\ =\sum_j \tr\left(\hat{U}^\dagger \hat{O}_M^{(j)} \hat{U} \ketbra{\psi_0} \right) \hat{O}_M^{(j)}
\label{eq:pheno1}
\end{eqnarray}

Since, $\hat{U}^\dagger \hat{O}_M^{(j)} \hat{U}$ can be calculated from the descriptors $\left(\left(\hat{U}^\dagger \hat{f}_{i_1} \hat{U}\dots , \hat{U}^\dagger \hat{f}_{i_M} \hat{U}\right), \ketbra{\psi_0}\right)$ and $\ketbra{\psi_0}$ is part of the ontic state, we can assign from our ontic local states the phenomenal local states through the epimorphism $\varphi$ given by:

\begin{eqnarray}
    \varphi\left(\left(\hat{U}^\dagger \hat{f}_{i_1} \hat{U}, \dots, \hat{U}^\dagger \hat{f}_{i_M} \hat{U}\right), \ketbra{\psi_0}\right)=\nonumber \\=\sum_j \tr\left(\hat{U}^\dagger \hat{O}_M^{(j)} \hat{U} \ketbra{\psi_0} \right) \hat{O}_M^{(j)}
    \label{eq:pheno}
\end{eqnarray}

The complete details of the decomposition we use are in the proof of theorem \ref{thm:1.5} in Appendix \ref{sec:appendix}. 

\subsection{Separability}
\label{subsec:separability}

Once we have a representation of the epimorphism $\varphi$ in terms of fermionic descriptors, we can discuss separability. Separability is the condition that gives the notion of locality to the ontic state-space construction. In the fermionic descriptor setting, we now discuss how we represent the ontic local states, how the join product operates, and how they relate to the local phenomenal states (the reduced density operators). \\

\subsubsection{Ontic projections}
\label{subsubsec:realproj}

We know that given a subset $I_A$ of the mode set $I_A \subseteq I$: $\{a_1,\dots,a_M\}\subseteq \{i_1,\dots,i_N\}$ induces an $M$ mode subsystem $A$ of the $N$ mode system $S$. Where $A$ is the fermionic system of $M$ modes generated by the mode set $I_A=\{i_1,\dots,i_M\}$. At the ontic state space in the {equivalence class formalism} to define the ontic projection mappings of the equivalence classes, one just defines $\pi^{\pazocal{R}}_A\left([U]_{AB}\right)=[U]_A$. It is straightforward to see that the following simple operation defines the ontic projection mappings in the descriptor representation of fermionic systems by using theorem \ref{thm:1}.  \\

\begin{defi}
If we have an ontic global state of the system $S$ represented by $\left(\left(\hat{U}^\dagger \hat{f}_{i_1} \hat{U},\dots,\hat{U}^\dagger \hat{f}_{i_N} \hat{U}\right),\ketbra{\psi_0}\right)$. Then, we define
\begin{eqnarray}
    \pi_A^\pazocal{R}\left(\left(\hat{U}^\dagger \hat{f}_{i_1} \hat{U}, \dots, \hat{U}^\dagger \hat{f}_{i_N} \hat{U}\right),\ketbra{\psi_0}\right)=\nonumber \\ =\left(\left(\hat{U}^\dagger \hat{f}_{a_1} \hat{U}, \dots, \hat{U}^\dagger \hat{f}_{a_M} \hat{U}\right),\ketbra{\psi_0}\right)
\end{eqnarray}
\end{defi}

\subsubsection{Joint ontic product}
\label{subsubsec:jointprod}

To define the ontic join product, we need to see that for any fermionic subsystem $A$ defined in terms of modes subsets, we can consider a bipartition of the global system $S$. We can consider the fermionic system $B$ of $N-M$ modes spanned by the set of modes $I_B=I\backslash I_A$. It is straightforward to see that $A$ and $B$ are disjoint systems and that the global system $S$ can be considered the join space of both subsystems.\\

Now, we are in a position to define ontic states of systems $A$ and $B$ to be compatible. Consider an ontic state of system $A$ represented with descriptors, $\left(\left(\hat{U}^\dagger \hat{f}_{a_1} \hat{U}, \dots, \hat{U}^\dagger \hat{f}_{a_M} \hat{U}\right),\ketbra{\psi}\right)$  and an ontic state of system $B$ represented with descriptors, $\left(\left(\hat{V}^\dagger \hat{f}_{b_1} \hat{V}, \dots, \hat{V}^\dagger \hat{f}_{b_{N-M}} \hat{V}\right),\ketbra{\eta}\right)$. Then, the state are said to be compatible if exists a global state of the system $S=A \cup B$, represented by descriptors as $\left(\left(\hat{W}^\dagger \hat{f}_{i_1} \hat{W}, \dots, \hat{W}^\dagger \hat{f}_{i_{N}} \hat{W}\right),\ketbra{\phi}\right)$ such that when projected to subsystems $A$ and $B$ equals the beforementioned states.\\

Note that this means that for two states to be compatible, they have to have the same Heisenberg state and that there must exist a unitary transformation $W$ such that $[W]_A=[U]_A$ and $[W]_B=[V]_B$. Then, it is straightforward for compatible states to see how the join product can be defined and operates in the descriptor representation. 

\begin{defi}
Consider two compatible local ontic states of disjoint subsystems $A$ and $B$, $\left(\left(\hat{U}^\dagger \hat{f}_{a_1} \hat{U}, \dots, \hat{U}^\dagger \hat{f}_{a_M} \hat{U}\right),\ketbra{\psi_0}\right)$ and $\left(\left(\hat{V}^\dagger \hat{f}_{b_1} \hat{V}, \dots, \hat{V}^\dagger \hat{f}_{b_{N-M}} \hat{V}\right),\ketbra{\psi_0}\right)$. Then it exists a unique global state  $\left(\left(\hat{W}^\dagger \hat{f}_{i_1} \hat{W}, \dots, \hat{W}^\dagger \hat{f}_{i_{N}} \hat{W}\right),\ketbra{\psi}\right)$ such that the definition of the join product of the compatible states can be defined as 

\begin{eqnarray}
   &\text{ \small $\left( \hat{U}^\dagger \left( \hat{f}_{a_1},\dots,\hat{f}_{a_M} \right) \hat{U},\ketbra{\psi_0}\right) \odot \left(\hat{V}^\dagger \left( \hat{f}_{b_{1}},\dots,\hat{f}_{b_{N-M}} \right)\hat{V},\ketbra{\psi_0}\right)=$}\nonumber \\&=\text{\small $\left(\left(\hat{W}^\dagger \hat{f}_{a_{1}} \hat{W},\dots,\hat{W}^\dagger \hat{f}_{b_N} \hat{W}\right),\ketbra{\psi_0}\right)$}
\end{eqnarray}
\end{defi}

 Thus, due to the uniqueness property of the definition, knowing the ontic states of the subsystems is the same as knowing the ontic state of the global system. The projection and join structure is the same for the fermionic case to the qubit case \cite{Raymond,Deutsch}. However, the critical result for the fermionic case is that for a global system where we know the unitaries that are applied, the projection and joining operations consist in separate and merge the descriptors of the relevant set of modes that conform to our subsystems of interest. \\

\subsection{Faithfulness of splitting operation}
\label{subsubsec:faithful}

Of course, we can repeat the bipartition process until we reach the point of individual modes forming subsystems. We see how the local and global states are on the same footing, which gives us strong notions to believe in their physicality given a particular partition of the global system in terms of subsystems. A particular excellent result that one can obtain from the representation of ontic states with descriptors is that splitting into subsystems is a faithful operation because the diagram in Figure \ref{fig:diagram} commutes.  Theorem \ref{thm:2} condenses the information in Figure \ref{fig:diagram}.

\begin{figure}
    \centering
    \includegraphics[width=0.4\textwidth]{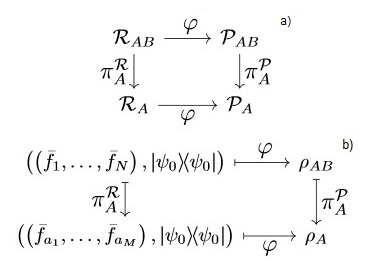}
    \label{fig:diagram}
    \caption{Commuting diagram that represents the actions of taking the projection into subsystems and the ontic-phenomenal epimorphism. Diagram a) represents the spaces, and the diagram b) represents the action of the mappings in the descriptor picture.  }
    \label{fig:diagram}
\end{figure}

\begin{theorem}
\label{thm:2}
Given a fermionic system with the mode set $I=\{1,\dots,N\}$. For any set of descriptors $(\hat{\bar{f}}_1,\dots,\hat{\bar{f}}_N)$ together with any Heisenberg state $\ketbra{\psi_0}$ and any non-empty subset of modes $J\subseteq I$. The associated diagram of Figure \ref{fig:diagram} commutes. In other words: 
\begin{eqnarray}
    \left(\pi^{\pazocal{P}}_J\circ \varphi\right)\left(\left(\hat{\bar{f}}_1,\dots,\hat{\bar{f}}_N\right),\ketbra{\psi_0}\right)=\nonumber \\ =\left(\varphi \circ \pi^{\pazocal{R}}_J\right)\left(\left(\hat{\bar{f}}_1,\dots,\hat{\bar{f}}_N\right),\ketbra{\psi_0}\right)
\end{eqnarray}
\end{theorem}

The proof of the theorem is algebraic. We apply the form of the epimorphism $\varphi$ for descriptors that we have on equation \ref{eq:pheno}. We also use the definition of ontic projection $\pi_A^{\pazocal{R}}$ that consists in forgetting the descriptors for the modes that are not in $A$. Furthermore, finally, we use the properties of $\pi_A^\pazocal{P}$ being the fermionic partial trace. Moreover, we use some algebraic properties of fermionic systems under the parity {superselection rule} derived in previous works \cite{Nicetu}.  The complete proof is in Appendix \ref{sec:appendix}.\\

The definitions and theorems allow us to claim that fermions also have as qubits a local-realistic structure. We have found explicitly a  neat representation of the ontic states using the widely known Heisenberg picture of quantum mechanics. Furthermore, we have seen that tracking the creation or annihilation operators is enough to have a complete description of the ontic states of the system. In the following section \ref{sec:discussion} we discuss the implications of our findings and the possible uses that this way of representing fermionic systems may have.  

\section{Discussion}
\label{sec:discussion}

Following the {equivalence class formalism}, this work shows how one can formally prove that fermionic systems are local-realistic. We have seen how a compact representation of the local structure can be given by extending the descriptor picture of qubit networks to fermionic modes. We have also seen that fermionic annihilation operators form a valid set of fermionic descriptors.  \\

Readers familiar with the Heisenberg picture in quantum field theory may wonder how the analysis we have performed differs from it. We are indeed just using the usual Heisenberg picture. We have seen that the most efficient way to use the Heisenberg picture is to track the fermionic annihilation operators since, from their evolution, we can retrieve the evolution of any observable of the theory. However, the interpretation of these objects by the {equivalence class formalism} differentiates the usual Heisenberg picture from our treatment. First, we have stressed the relevance and necessity of having local objects associated with each fermionic mode, which are the subsystems in our theory. And second, we propose to interpret the dynamical evolution of the annihilation operators as a representation of the equivalence classes of {parity superselected} unitaries that correspond to the ontic states in our system. This consistent interpretation is a novel element that introduces a possible new angle on the meaning of states and dynamics in fermionic systems. \\

Marletto et al.  \cite{chiara} have analysed how a phase in a fermionic Mach-Zender interferometer is acquired locally by using the Heisenberg picture and tracking some relevant observables. This paper is a generalisation of that work, using the new {construction} of the {equivalence class formalism}. We specify the notion of a local element of reality in terms of local ontic states, and we can then describe any fermionic physical process with a local description. \\   

We emphasise, as also pointed out by Chiribella \cite{Chiribella}, that the lattice of systems of a theory is an extra layer of the structure that the fermionic algebra alone does not determine. More concretely, in the fermionic case, the lattice of systems is determined by which notion of modes one chooses. Moreover, under that partition, we can assign the ontic local states in terms of subsystems of the theory. However, the choice of modes is not unique. We could perfectly define a different set of modes by unitary conjugation to the chosen one that would define a different partitioning of the global system in terms of subsystems. Furthermore, in this other partition, we would be able to identify the ontic local states. Therefore, even though we can uniquely identify the local elements of reality given a notion of locality, the locality is not unique for a group of physical transformations. \\

For a set of local operators to be a suitable set of descriptors, we need that the evolution of such operators suffices to determine the evolution of any observable. One can wonder if there is a straightforward way to understand why the annihilation operators are suitable descriptors. There is. We first observe that fermionic observables are fermionic operators, which form an algebra. We now notice that in this algebra, in particular, the $2N$ creation and annihilation operators are generators of the algebra of fermionic operators. And since $\hat{U}^\dagger(\alpha \hat{A}+\hat{B})\hat{U}=\alpha U^\dagger(A)U+U^\dagger(B)U$ and $\hat{U}^\dagger(\hat{A} \hat{B})\hat{U}=\hat{U}^\dagger(\hat{A})\hat{U} \hat{U}^\dagger(\hat{B})\hat{U}$, tracking the generators of the algebra is enough to track any element of it. Using the fact that $\left(\hat{f}\right)^\dagger=\hat{f}^\dagger$ we can claim that tracking $\hat{f}$ is enough to track $\hat{f}^\dagger$, so it is enough to track the annihilation operators.

This observation is useful in understanding why in the $N$ qubit network case, the descriptors can be the $\hat{\sigma}^x_j,\hat{\sigma}^z_j$ operators for each qubit $j$. These $\sigma^x_j,\sigma^z_j$ are the generators of the $N$ qubit network operator algebra. Following the same spirit as in this article, we can now also use the $\dagger$ operation properties. It is straightforward to see that if one defines $\hat{q}_j=\frac{1}{2}\left(\hat{\sigma}^x_j+i\hat{\sigma}^y_j\right)$, then $\hat{q}_j, \hat{q}_j^\dagger$ are generators of the $N$ qubit algebra (see in Appendix \ref{sec:appendix}). We can see that since $\left(\hat{q}_j\right)^\dagger=\hat{q}_j^\dagger$, then is enough to track all the $\hat{q}_j$ to track any $N$ qubit observable. Therefore we only need to track one operator for each qubit, with the caveat of that operator not being an observable since it is a non-hermitian operator.\\

These operators $\hat{q}_j$ behave like annihilation operators. They satisfy the condition that $\hat{q}_j\ket{0}\dots\ket{0}=0$, $\hat{q}_j^2=0$, $\{\hat{q}_j,\hat{q}_j^\dagger\}=\hat{\mathbb{I}}$ and $[\hat{q}_j,\hat{q}_i]=0$. At the individual qubit/mode level, the anticommutation relations are the same as for fermionic modes, but operators of different modes/qubits commute. This realisation is not related to the Jordan-Wigner transformation; it is just a property that arises from the defining commutation/anticommutation relations of generators of the qubit network algebra. \\

We can now postulate what the descriptors will be for any quantum system of particles. For the bosonic case, we can see that the direct candidate for bosonic descriptors in a system of $N$ bosonic modes is the set of bosonic annihilation operators $\hat{b}_j$. We know that the set of $\hat{b}_j, \hat{b}_j^\dagger$ are the generators of the bosonic operator algebra. We can simplify further, by tracking only the set of $\hat{b}_j$ -- this is enough to track any bosonic observable. In our future research, we would like to use the notion of algebra generators to study the locality and descriptors of anyonic systems \cite{stevesimon}, where the identifications are not so clear due to the lack of field theory creation and annihilation operators. Similarly, we would like to expand on these concepts focusing on well-known quantum algebras such as the Virasoro algebra used in string theory to find a compact local description of such systems.    \\

A confusing question can arise in our main fermionic case. Due to the parity {superselection rule}, $\hat{f}_i$ are not fermionic observables, nor fermionic unitaries, nor a linear combination of them; how can the collection of them together with $\ketbra{\psi_0}$ be the ontic state of the {parity super selected} fermionic system? We answer this by regarding the set of fermionic annihilation operators and $\ketbra{\psi_0}$ not as the ontic state of the fermionic system; but as a representation of the ontic state. We use here representation as a broad term, not the precise meaning used in group theory. The ontic states are the equivalence classes of the {parity superselected} unitaries that we have described with the equivalence relation. Note that only the {parity superselected} unitaries (real, physically allowed) are required to define such equivalence classes as the ontic states. We use the descriptors and the Heisenberg picture to have a compact and convenient way of representing the ontic states of the {parity superselected} unitaries' equivalence classes and their structure. \\

Finally, in the fermionic and the general {equivalence class formalism} case, only the set of physically allowed transformations is necessary to define the ontic states of a physical system. Thus, we think these results suggest that we should study the transformation structures by themselves to see what range of structures and phenomena can offer. Identifying which transformations are physically possible seems crucial in understanding the constitutive properties and the locality features of a physical system. Further future work could involve extending the {equivalence class formalism} to non-reversible dynamics so that theories such as the objective collapse model \cite{objectivecollapse} could be studied in detail.

\section*{Acknowledgements}
We thank David Deutsch, Paul Raymond-Robichaud and Charles Alexandre B\'edard for fruitful discussions. NTV acknowledges financial support from 'la Caixa' Foundation (ID 100010434, LCF/BQ/EU18/11650048). CM and VV thank the John Templeton Foundation and the Eutopia Foundation. VV's research is supported by the National Research Foundation and the Ministry of Education in Singapore and administered by Centre for Quantum Technologies, National University of Singapore. This publication was made possible through the support of the ID 61466 grant from the John Templeton Foundation, as part of the The Quantum Information Structure of Spacetime (QISS) Project (qiss.fr). The opinions expressed in this publication are those of the authors and do not necessarily reflect the views of the John Templeton Foundation.

\subsection*{Conflict of interest}
The authors have no conflicts to disclose.

\section*{Data Availability Statement}
Data sharing is not applicable to this article as no new data were created or analyzed in this study.

\onecolumngrid
\appendix

\section{{Proofs and mathematical details}}
\label{sec:appendix}

In this section, we include mathematical details of the paper and laborious proofs of statements made in the main article. \\

\begin{theorem*} \ref{thm:1})
Given a fermionic theory with a set of modes  $I=\{i_1,\dots,i_N\}$. Given the equivalence relation on the group of {parity superselected} unitaries of the theory for each mode $i_j\in I$ given by $\hat{U}\sim_{i_j} \hat{V}$ iff $\hat{U}=\hat{W}_{I\backslash\{i_j\}} \cdot \hat{V}$, then 

\begin{eqnarray*}
    \hat{U} \sim_{i_j} \hat{V} \qquad  \Longleftrightarrow \qquad U^\dagger \cdot \hat{f}_{i_j} \cdot \hat{U}=\hat{V}^\dagger \cdot \hat{f}_{i_j} \cdot \hat{V}
\end{eqnarray*}
Thus, $[\hat{U}]_{i_j}=\{\hat{V}\in \pazocal{T}| \hat{U}^\dagger \cdot \hat{f}_{i_j} \cdot \hat{U}=\hat{V}^\dagger \cdot \hat{f}_{i_j} \cdot \hat{V} \}$
\end{theorem*}

\begin{proof}
The last statement follows directly from the definition of an equivalence class, so the equation that needs to be proven is eq. \ref{eq:equiv}: \\ 
"$\Rightarrow$": $\hat{U}\sim_{I_j} \hat{V}$ implies $\hat{U}=\hat{W}_{I\backslash \{i_j\}} \cdot \hat{V}$ for some $\hat{W}_{I\backslash \{i_j\}}$ being a {parity superselected} unitary local on the set of modes that excludes the mode $i_j$. Thus, since $\hat{W}_{I\backslash \{i_j\}}$ is an even operator that does not contain any terms involving $\hat{f}_{i_j},\hat{f}_{i_j}^\dagger$ is straightforward to check that $[\hat{W}_{I\backslash \{i_j\}},\hat{f}_{i_j}]=0$. Thus, then: $\hat{U}^\dagger \cdot \hat{f}_{i_j} \cdot \hat{U}=\hat{V}^\dagger \cdot \hat{W}_{I\backslash \{i_j\}}^\dagger \cdot \hat{f}_{i_j}\cdot \hat{W}_{I\backslash \{i_j\}} \cdot \hat{V}= \hat{V}^\dagger \cdot \hat{f}_{i_j} \cdot \hat{W}_{I\backslash \{i_j\}}^\dagger \cdot \hat{W}_{I\backslash \{i_j\}} \cdot \hat{V}=\hat{V}^\dagger \cdot \hat{f}_{i_j} \cdot \hat{V}$. \\
"$\Leftarrow$": We have that $\hat{U}^\dagger \cdot \hat{f}_{i_j} \cdot \hat{U}=\hat{V}^\dagger \cdot \hat{f}_{i_j} \cdot \hat{V}$, and to see that $\hat{U}\sim_{i_j} \hat{V}$ we need to see that $\hat{U}=\hat{W}_{I\backslash \{i_j\}} \cdot \hat{V}$. Or, equivalently, since we have a group structure, where transformations are unitaries, proving that $\hat{U}\cdot \hat{V}^\dagger=\hat{W}_{I\backslash \{i_j\}}$ is enough to proof that $\hat{U}\sim_{i_j}\hat{V}$. From $\hat{U}^\dagger \cdot \hat{f}_{i_j} \cdot \hat{U}=\hat{V}^\dagger \cdot \hat{f}_{i_j} \cdot \hat{V}$ is straightforward to deduce that then $ \hat{f}_{i_j} \cdot (\hat{U} \cdot \hat{V}^\dagger)=(\hat{U}\cdot \hat{V}^\dagger) \cdot \hat{f}_{i_j}$. Naming $\hat{U}\cdot \hat{V}^\dagger=\hat{W}$, noticing $\hat{W}$ is a {parity superselected} unitary and taking the dagger of the previous equation we have that the two following equalities hold: 
\begin{eqnarray}
    \hat{W}\cdot  \hat{f}_{i_j}=\hat{f}_{i_j}\cdot \hat{W} \qquad \qquad  \hat{W}\cdot  \hat{f}^\dagger_{i_j}=\hat{f}^\dagger_{i_j}\cdot \hat{W}\label{eq:conditions}
\end{eqnarray}
Moreover, now since $W$ is a priori a general {parity superselected} unitary, is not dificult to see that we can decompose it as: $\hat{W}=\hat{O}_0+\hat{f}_{i_j} \hat{O}_1 +\hat{f}^\dagger_{i_j} \hat{O}_2 + \hat{f}_{i_j} \hat{f}^\dagger_{i_j} \hat{O}_3$. Where $\hat{O}_0,\hat{O}_3,\hat{O}_1,\hat{O}_2$ are local operators on the set of modes $I\backslash \{i_j\}$, where $O_0,O_3$ are even operators and $\hat{O}_1,\hat{O}_2$ are odd operators (since the operator $\hat{W}$ is an even operator). Using this decomposition of $\hat{W}$ in the first condition of eq. \ref{eq:conditions} and commuting/anticommuting the $\hat{f}_{i_j},\hat{f}_{i_j}^\dagger$ terms with the $\hat{O}_k$ operators we obtain that: 
\begin{eqnarray*}
     \hat{f}_{i_j}( \hat{O}_0+\hat{O}_3) + \hat{f}_{i_j} \hat{f}_{i_j}^\dagger \hat{O}_2-\hat{O}_2 =\hat{f}_{i_j} \hat{O}_0+\hat{f}_{i_j}\hat{f}_{i_j}^\dagger \hat{O}_2
\end{eqnarray*} 
Is not difficult to see that this implies that $\hat{O}_2=\hat{0}$ and $\hat{O}_3=\hat{0}$. Then, using that $\hat{W}=\hat{O}_0+\hat{f}_{i_j} \hat{O}_1$ and replacing in the second condition of eq. \ref{eq:conditions} we obtain: 
\begin{eqnarray*}
  \hat{f}_{i_j}^\dagger \hat{O}_0+\hat{f}_{i_j} \hat{f}_{i_j}^\dagger \hat{O}_1= \hat{f}_{i_j}^\dagger \hat{O}_0+\hat{O}_1-\hat{f}_{i_j}^\dagger \hat{f}_{i_j} \hat{O}_1 
\end{eqnarray*}
That is also easy to see that implies that $\hat{O}_1=\hat{0}$. Thus, we have seen that the conditions imply that $\hat{W}=\hat{O}_0$, thus being a local {parity superselected} unitary on the set of modes $I\backslash\{i_j\}$, thus allowing us to name $\hat{W}=\hat{W}_{I\backslash\{i_j\}}$, and therefore we have proven that $\hat{U}\sim_{i_j} \hat{V}$.  
\end{proof}

\begin{theorem*}\ref{thm:1.5})
Using the complete set of descriptors $\left(\hat{\bar{f}}_1,\dots,\hat{\bar{f}}_N\right)$ is possible to find uniquely the $\hat{U}$ that acts on the $N$ mode system such that $\left(\hat{\bar{f}}_1,\dots,\hat{\bar{f}}_N\right)=\left(\hat{U}^\dagger \hat{f}_1 \hat{U},\dots,\hat{U}^\dagger \hat{f}_N \hat{U}\right)$. 
\end{theorem*}

\begin{proof}
First, note that the unitary conjugation action is up to a global phase. Keeping this in mind, considering the unitary $\hat{U}$ as an operator, we can regard it as a vector on the vector space of operators. We can consider in the operator algebra an orthonormal basis, where the scalar product between two operators $\hat{A},\hat{B}$ is given by $\tr(\hat{A}^\dagger \hat{B})$. It is straightforward to see that for the fermionic operators, if one labels the orthonormal Fock basis as $\{\ket{k}\}_{k=1}^{2^N}$, then $\ketbra{k}{l}$ is an orthonormal basis of the operator vector space with scalar product given by $\tr{\hat{A}^\dagger \hat{B}}$. \\

Considering this, now $\hat{U}$ can be written as $\hat{U}=\sum_{k,l=1}^{2^N} \tr\left(\hat{U} \ketbra{k}{l}\right) \ketbra{k}{l}$. So, if we know $\tr\left(\hat{U} \ketbra{k}{l}\right)$ we know the unitary. $\ketbra{k}{l}$ is a product of creation and annihilation operators, since $\ket{k}=\hat{f}^\dagger_{i_1} \dots \hat{f}^\dagger_{i_n} \ket{\Omega}$ and $\ketbra{\Omega}=\hat{f}_N \dots \hat{f}_1 \hat{f}^\dagger_1 \dots \hat{f}^\dagger_N$. Using this fact, having at our disposal $\left(\hat{f}_1,\dots,\hat{f}_N\right)$ and  $\left(\hat{\bar{f}}_1,\dots,\hat{\bar{f}}_N\right)$ we can construct $\ketbra{k}{l}$ and $\ketbra{\bar{k}}{\bar{l}}$, where to construct the second we have replaced the $\hat{f}_i, \hat{f}_i^\dagger$ in the decomposition of $\ketbra{k}{l}$ by $\hat{\bar{f}}_i,\hat{\bar{f}}_i^\dagger$. We can see easily that $\ketbra{\bar{k}}{\bar{l}}=\hat{U}^\dagger \ketbra{k}{l} \hat{U}$. \\

We take the scalar product of any two of these objects. In other words, consider $\tr\left(\ketbra{\bar{k}}{\bar{l}} \ketbra{m}{n}\right)=\tr\left(\hat{U}^\dagger\ketbra{k}{l}\hat{U} \ketbra{m}{n}\right)$. Now, if we insert the decomposition of $\hat{U}$ found above and we use the linearity properties of the trace, we obtain that 
\begin{eqnarray*}
   \tr\left(\ketbra{\bar{k}}{\bar{l}} \ketbra{m}{n}\right) =\text{\small $\sum_{o,p,q,r=1}^{2^N} \tr\left(\hat{U}^\dagger \ketbra{o}{p}\right) \tr\left(\hat{U} \ketbra{q}{r}\right) \tr\left(\ketbra{p}{o} \cdot \ketbra{k}{l} \cdot \ketbra{r}{q} \cdot \ketbra{m}{n}\right)$}= \sum_{o,p,q,r=1}^{2^N} \tr\left(\hat{U}^\dagger \ketbra{o}{p}\right) \tr\left(\hat{U} \ketbra{q}{r}\right) \delta_{o k} \delta_{l r} \delta_{q m} \delta_{n p}=\\ =\tr\left(\hat{U}^\dagger \ketbra{k}{n}\right) \tr\left(\hat{U} \ketbra{m}{l}\right) 
\end{eqnarray*}
where we have used the orthonormality of the Fock basis and the properties of the Kronecker delta. Using the cyclic properties of the trace and complex conjugation we obtain that $\tr\left(\hat{U}^\dagger \ketbra{k}{n}\right) \tr\left(\hat{U} \ketbra{m}{l}\right)=\tr\left(\hat{U} \ketbra{n}{k}\right)^* \tr\left(\hat{U} \ketbra{m}{l}\right)$. The question that now arises is that if knowing all the values of $\tr\left(\ketbra{\bar{k}}{\bar{l}} \ketbra{m}{n}\right)$ (that we can obtain since we only use $\left(f_1,\dots,f_N\right)$ and  $\left(\bar{f}_1,\dots,\bar{f}_N\right)$), we can retrieve $\tr\left(\hat{U} \ketbra{m}{l}\right)$. We can. We need to notice that $\tr\left(\hat{U} \ketbra{m}{l}\right)$ is a complex number so knowing its polar form is enough. We see that $\tr\left(\ketbra{\bar{l}}{\bar{l}} \ketbra{m}{m}\right)=\left|\tr\left(\hat{U} \ketbra{m}{l}\right)\right|^2$. Thus, we obtain the modulus of the complex number.
We can see now that $\tr\left(\hat{U} \ketbra{m}{l}\right)=\sqrt{\tr\left(\ketbra{\bar{l}}{\bar{l}} \ketbra{m}{m}\right)} e^{i \phi_{m,l}}$. So only the phases are up to determine. Here is where the issue of the overall phase freedom intervenes. Since $\hat{U}$ is unitary, we know they must exist $m_0,l_0$ such that $\left|\tr\left(\hat{U} \ketbra{m_0}{l_0}\right)\right|^2=\tr\left(\ketbra{\bar{l_0}}{\bar{l_0}} \ketbra{m_0}{m_0}\right) >0$. We have the freedom to fix the phase $\phi_{m_0,l_0}=0$ due to the overall phase redundancy. 
In other words, we could always choose a global phase to cancel the phase $\phi_{m_0,l_0}$ so it is set to $0$. Now, we can see that if we consider $\tr\left(\ketbra{\bar{l_0}}{\bar{l}} \ketbra{m}{m_0}\right)=\tr\left(\hat{U} \ketbra{m_0}{l_0}\right)^* \tr\left(\hat{U} \ketbra{m}{l}\right)=\sqrt{\tr\left(\ketbra{\bar{l_0}}{\bar{l_0}} \ketbra{m_0}{m_0}\right)} \sqrt{\tr\left(\ketbra{\bar{l}}{\bar{l}} \ketbra{m}{m}\right)}  e^{i \phi_{m l}}$. Thus, we obtain that 
\begin{eqnarray*}
    \tr\left(\hat{U} \ketbra{m}{l}\right)=\frac{\tr\left(\ketbra{\bar{l_0}}{\bar{l}} \ketbra{m}{m_0}\right)}{\sqrt{\tr\left(\ketbra{\bar{l_0}}{\bar{l_0}} \ketbra{m_0}{m_0}\right)}}
\end{eqnarray*}
Therefore, indeed we can retrieve the unitaries that conjugate $\left(\hat{\bar{f}}_1,\dots,\hat{\bar{f}}_N\right)$ to $\left(\hat{f}_1,\dots,\hat{f}_N\right)$. 
\end{proof}

\begin{theorem*}{\ref{thm:2})}
Given a fermionic system with the mode set $I=\{i_1,\dots,i_N\}$. For any set of descriptors $(\hat{\bar{f}}_1,\dots,\hat{\bar{f}}_N)$ together with any Heisenberg state $\ketbra{\psi_0}$ and any non-empty subset of modes $J\subseteq I$. The associated diagram of Figure \ref{fig:diagram} commutes. In other words: 
\begin{eqnarray*}
    \text{\small $\left(\pi^{\pazocal{P}}_J\circ \varphi\right)\left(\left(\hat{\bar{f}}_1,\dots,\hat{\bar{f}}_N\right),\ketbra{\psi_0}\right)=\left(\varphi \circ \pi^{\pazocal{R}}_J\right)\left(\left(\hat{\bar{f}}_1,\dots,\hat{\bar{f}}_N\right),\ketbra{\psi_0}\right)$}
\end{eqnarray*}
\end{theorem*}

\begin{proof}
We begin by expanding the right hand side of the equation, by applying the definition of the ontic projection operator, obtaining: 
\begin{eqnarray*}
    \left(\varphi \circ \pi^{\pazocal{R}}_J\right)\left(\left(\hat{\bar{f}}_1,\dots,\hat{\bar{f}}_N\right),\ketbra{\psi_0}\right)=\varphi\left(\left(\hat{\bar{f}}_{j_1},\dots,\hat{\bar{f}}_{j_n}\right),\ketbra{\psi_0}\right)
\end{eqnarray*}
Applying now the definition of $\varphi$ of Eq. \ref{eq:pheno}, we obtain: 
\begin{eqnarray}
   \varphi\left(\left(\hat{\bar{f}}_{j_1},\dots,\hat{\bar{f}}_{j_n}\right),\ketbra{\psi_0}\right)=  \sum_{\vec{p},\vec{l}} \tr\left(\hat{U}^\dagger \left(\hat{f}_{l_1}^\dagger \dots \hat{f}_{l_m}^\dagger \ketbra{\Omega} \hat{f}_{p_r} \dots \hat{f}_{p_1}\right) \hat{U} \ketbra{\psi_0}\right)   \hat{f}_{l_1}^\dagger \dots \hat{f}_{l_m}^\dagger \ketbra{\Omega} \hat{f}_{p_r} \dots \hat{f}_{p_1} 
\end{eqnarray}
where we have chosen as orthonormal basis for the operator space of the fermionic system with modes $j_1,\dots,j_n$ the basis given by $\{\hat{f}_{l_1}^\dagger \dots \hat{f}_{l_m}^\dagger \ketbra{\Omega} \hat{f}_{p_r} \dots \hat{f}_{p_1}\}$. Where, $l_1 < \dots < l_m$, $p_1 < \dots < p_r$ with $l_1,\dots,l_m,p_1,\dots,p_r \in J$, where $\ket{\Omega}$ is the vacuum for the set of modes $I$, and thus also $J$.  \\

We now turn to expand the left-hand side of the initial equation by applying the definition of the epimorphism and choosing the same basis as before, but for the global set of modes \{1,\dots, N\}, we obtain: 
\begin{eqnarray*}
    \left(\pi^{\pazocal{P}}_J\circ \varphi\right)\left(\left(\hat{\bar{f}}_1,\dots,\hat{\bar{f}}_N\right),\ketbra{\psi_0}\right)=\pi^{\pazocal{P}}_J \left( \sum_{\vec{p},\vec{l}} \tr\left(\hat{U}^\dagger \left(\hat{f}_{l_1}^\dagger \dots \hat{f}_{l_m}^\dagger \ketbra{\Omega} \hat{f}_{p_r} \dots \hat{f}_{p_1}\right) \hat{U} \ketbra{\psi_0}\right)   \hat{f}_{l_1}^\dagger \dots \hat{f}_{l_m}^\dagger \ketbra{\Omega} \hat{f}_{p_r} \dots \hat{f}_{p_1}  \right)
\end{eqnarray*}

The main difference with the expanded version of the right hand side is the values of the vectors $\vec{l},\vec{p}$, in the first expression they can have length up to $n$ with the values being elements of $J$, while in the second case they can have lengths up to $N$ with the values being elements of $I$.\\

Now, lets take the fermionic partial trace. In order to ease up on notation, lets reorder our set of modes such that $J$ correspond to the first $n$ elements of our set of modes $I$. First, lets trivially observe that if $J=I$ the equality holds, so, now we consider $J$ a strict subset of $I$. Using the reordering we can see how the expression of the left hand side becomes 
\begin{eqnarray*}
   \pi^{\pazocal{P}}_J \left( \sum_{\vec{p},\vec{l},\vec{u},\vec{v}} \tr\left(\hat{U}^\dagger \left(\hat{f}_{l_1}^\dagger \dots \hat{f}_{l_m}^\dagger \hat{f}_{u_1}^\dagger \dots \hat{f}_{u_s}^\dagger  \ketbra{\Omega} \hat{f}_{v_t} \dots \hat{f}_{v_1} \hat{f}_{p_r} \dots \hat{f}_{p_1}\right) \hat{U} \ketbra{\psi_0}\right)   \hat{f}_{l_1}^\dagger \dots \hat{f}_{l_m}^\dagger  \hat{f}_{u_1}^\dagger \dots \hat{f}_{u_s}^\dagger \ketbra{\Omega}  \hat{f}_{v_t} \dots \hat{f}_{v_1} \hat{f}_{p_r} \dots \hat{f}_{p_1}  \right)
\end{eqnarray*}
 
 where $\vec{p},\vec{l}$ are vectors of elements of $J$ up to length $n$ and the vectors $\vec{u},\vec{v}$ are vectors of elements of $I\backslash J$ up to length $N-n$. Once we have the expression of this form, we first notice that $\rho$ will be a {parity superselected} state, since $\rho_0$ is and the unitary is also, and the trace of an odd operator is always zero. With this notion, we can apply the properties of the fermionic partial trace \cite{Nicetu} where we need $\vec{u}=\vec{v}$ for the component not to vanish. Thus, we obtain that the left hand side becomes: 
 
 \begin{eqnarray*}
   \sum_{\vec{p},\vec{l},\vec{u}} \tr\left(\hat{U}^\dagger \left(\hat{f}_{l_1}^\dagger \dots \hat{f}_{l_m}^\dagger \hat{f}_{u_1}^\dagger \dots \hat{f}_{u_s}^\dagger  \ketbra{\Omega} \hat{f}_{u_s} \dots \hat{f}_{u_1} \hat{f}_{p_r} \dots \hat{f}_{p_1}\right) \hat{U} \ketbra{\psi_0}\right)   \hat{f}_{l_1}^\dagger \dots \hat{f}_{l_m}^\dagger \ketbra{\Omega}  \hat{f}_{p_r} \dots \hat{f}_{p_1} 
\end{eqnarray*}
 
 Moreover, we can focus now only on the coefficients to see that we can pull the sum over the vector $\vec{u}$ inside. By using the cyclic property of the trace and then using its linearity, we obtain that the coefficients of the left-hand side are: 
 
 \begin{eqnarray*}
     \text{\small$\sum_{\vec{u}}\tr\left( \left(\hat{f}_{l_1}^\dagger \dots \hat{f}_{l_m}^\dagger \hat{f}_{u_1}^\dagger \dots \hat{f}_{u_s}^\dagger  \ketbra{\Omega} \hat{f}_{u_s} \dots \hat{f}_{u_1} \hat{f}_{p_r} \dots \hat{f}_{p_1}\right) \hat{U} \ketbra{\psi_0} \hat{U}^\dagger\right)= \tr\left( \left(\hat{f}_{l_1}^\dagger \dots \hat{f}_{l_m}^\dagger \left( \sum_{\vec{u}} \hat{f}_{u_1}^\dagger \dots \hat{f}_{u_s}^\dagger  \ketbra{\Omega} \hat{f}_{u_s} \dots \hat{f}_{u_1} \right) \hat{f}_{p_r} \dots \hat{f}_{p_1}\right) \hat{U} \ketbra{\psi_0} \hat{U}^\dagger\right)$}
 \end{eqnarray*}
 
 Now, it can be used the fact that $\hat{f}_{l_1}^\dagger \dots \hat{f}_{l_m}^\dagger \hat{f}_{u_1}^\dagger \dots \hat{f}_{u_s}^\dagger \ket{\Omega}$ can be written as $\ket{l}\wedge \ket{u}$. Thus, obtaining that the coefficients of the left hand side can be written as: 
 
 \begin{eqnarray*}
    \tr\left( \left(\hat{f}_{l_1}^\dagger \dots \hat{f}_{l_m}^\dagger \left( \sum_{\vec{u}} \hat{f}_{u_1}^\dagger \dots \hat{f}_{u_s}^\dagger  \ketbra{\Omega} \hat{f}_{u_s} \dots \hat{f}_{u_1} \right) \hat{f}_{p_r} \dots \hat{f}_{p_1}\right) \hat{U} \ketbra{\psi_0} \hat{U}^\dagger\right)= \tr\left( \left(\ketbra{l}{p} \wedge \left(\sum_{u} \ketbra{u} \right) \right)\hat{U} \ketbra{\psi_0} \hat{U}^\dagger\right)
 \end{eqnarray*}
 
 we have that $\sum_{u} \ketbra{u} =\mathbb{I}$ and $\hat{O}\wedge \mathbb{I}=\hat{O}$, thus we obtain indeed that the left hand side is $\sum_{\vec{p},\vec{l}} \tr\left(\hat{U}^\dagger \left(\hat{f}_{l_1}^\dagger \dots \hat{f}_{l_m}^\dagger  \ketbra{\Omega} \hat{f}_{p_r} \dots \hat{f}_{p_1}\right) \hat{U} \ketbra{\psi_0}\right)$ $\hat{f}_{l_1}^\dagger \dots \hat{f}_{l_m}^\dagger \ketbra{\Omega}  \hat{f}_{p_r} \dots \hat{f}_{p_1}$. Precisely as the right-hand side, proving the theorem. 
 
\end{proof}

\begin{prop}
If one defines $\hat{q}_j=\frac{1}{2}\left(\hat{\sigma}^x_j+i\hat{\sigma}^y_j\right)$, then $\hat{q}_j, \hat{q}_j^\dagger$ are generators of the $N$ qubit algebra.
\end{prop}

\begin{proof}
It is well established that $\mathbb{I}_j,\hat{\sigma}^x_j,\hat{\sigma}_j^y,\hat{\sigma}_j^z$ are generators of the $N$ qubit algebra. Moreover, since $\left(\hat{\sigma}_j^x\right)^2=\mathbb{I}_j$ and $-i \hat{\sigma}_j^x \hat{\sigma}_j^y =\hat{\sigma}_j^z$ we can say that $\hat{\sigma}_j^x,\hat{\sigma}^y_j$ are generators of the $N$ qubit algebra. We will see how $\hat{q}_j,\hat{q}_j^\dagger$ are enough to generate such operators that generate the algebra; thus becoming generators. We first note that $\hat{q}_j+\hat{q}_j^\dagger=\hat{\sigma}^x_j$. And finally, $-i\left(\hat{q}_j-\hat{q}_j^\dagger\right)=\hat{\sigma}^y_j$.
\end{proof}

\twocolumngrid


\end{document}